\documentclass{sig-alternate-05-2015}
\usepackage[latin9]{inputenc}
\usepackage{url}
\usepackage{color}
\usepackage{soul}
\usepackage{booktabs}
\usepackage{graphicx}
\usepackage{multirow}
\usepackage{citesort}
\usepackage{mathrsfs}
\usepackage{amssymb}
\usepackage{amsbsy}
\usepackage{citesort}
\usepackage{listings}

\global\long\def\hb{\boldsymbol{h}}
\global\long\def\ib{\boldsymbol{i}}
\global\long\def\fb{\boldsymbol{f}}

\global\long\def\ob{\boldsymbol{o}}

\global\long\def\cb{\boldsymbol{c}}
\global\long\def\bb{\boldsymbol{b}}

\global\long\def\wb{\boldsymbol{w}}

\global\long\def\hb{\boldsymbol{h}}

\global\long\def\bb{\boldsymbol{b}}
\global\long\def\cb{\boldsymbol{c}}

\global\long\def\fb{\boldsymbol{f}}
\global\long\def\ib{\boldsymbol{i}}
\global\long\def\thetab{\boldsymbol{\theta}}

\usepackage{microtype}

\begin{document}


\doi{XXXXX}

\isbn{XXXXX}

\title{A deep language model for software code}

\numberofauthors{3} 
%
\author{
	%
	%
	\alignauthor
	Hoa Khanh Dam\\
	\affaddr{University of Wollongong \\Australia}\\
	\email{hoa@uow.edu.au}
	\alignauthor
	Truyen Tran\\
	\affaddr{Deakin University \\ Australia}\\
	\email{truyen.tran@deakin.edu.au}
	\alignauthor
	Trang Pham\\
	\affaddr{Deakin University \\ Australia}\\
	\email{phtra@deakin.edu.au}
}

\maketitle
\begin{abstract}
Existing language models such as n-grams for software code often fail to capture a long context where dependent code elements scatter far apart. In this paper, we propose a novel approach to build a language model for software code to address this particular issue. Our language model, partly inspired by human memory, is built upon the powerful deep learning-based Long Short Term Memory architecture that is capable of learning long-term dependencies which occur frequently in software code. Results from our intrinsic evaluation on a corpus of Java projects have demonstrated the effectiveness of our language model. This work contributes to realizing our vision for DeepSoft, an end-to-end, generic deep learning-based framework for modeling software and its development process.
 \end{abstract}

%
%

%
%



\section{Introduction}

Code is arguably the most important software artifact. Decades of research have been dedicated to understanding and modeling software code better. A useful representation of software code will support many important software engineering (SE) tasks such as code suggestion, code search, defect prediction, bug localization, test case generation, code patch generation, and even developer modeling. While high-level representations such as an abstract syntax trees or human-engineered features (e.g., lines of code, code complexity metrics, etc.) are useful for a certain analysis, they do not reveal the semantics hidden deeply in source code. On the other hand, low-level representations such as concrete syntax trees do not scale to the size of today's software where programs on the magnitude of a million lines of code are getting more common.

The past few years have witnessed the rise of Natural Language Processing (NLP) with many ground-breaking results in various applications such as speech recognition and machine translation. The success of NLP is rooted in its statistical, corpus-based methods and the availability of increasingly large corpora of text. Software code is also written in an (artificial) language, and has thus some characteristics that are the same  and different with natural language text:

\begin{enumerate}
  \item Repetitiveness: code fragments tend to reoccur very often \cite{Gabel:2010:SUS}. For example, loops such as \texttt{for (int i = 0; i < n; i++)} or common statements such as \texttt{System.out.println(...)} occur in many source files.

  \item Localness: source code is highly localized, i.e. there is a certain distinct repetitive patterns in a local context \cite{Tu:2014:LS}. For example, in a given source file, the pattern \texttt{for (int size} appears more often than the global pattern \texttt{for (int i}.

  \item Rich and explicit structural information: software code has highly complex structural information (which is different from natural language text). There are for example nested loops and inheritance hierarchies occurring frequently in source code.

  \item Long-term dependencies: a code element may depends on other code elements which are not immediately before it. For example, pairs of code tokens that are required to appear together due to programming language specification (e.g., try and catch in Java) or due to API usage specification (e.g., file open and close).

\end{enumerate}

Recent efforts have developed NLP-inspired language models for software code to realize some of those characteristics. They built a very large code corpus (by collecting publicly available source code from open source projects) and formed the vocabulary of a language. This vocabulary $\mathscr{V}$ consists of $N$ unique code tokens used across a substantial number of software projects. Given a code sequence $s=\langle w_{1},w_{2},...,w_{k}\rangle$ where $w_t$ is a code token (e.g., $s=\langle$\texttt{for, (, int, i, =, 0, i, ++,...}$\rangle$), a language model estimates the probability distribution $P(s)$ of this code sequence as follows.
\[
P(s)=P(w_{1})\prod_{t=2}^{k}P\left(w_{t}\mid\wb_{1:t-1}\right)
\]
where $\wb_{1:t-1}=(w_{1},w_{2},...,w_{t-1})$ is the historical \emph{context} used to estimate the probability of the next code token $w_t$.


The most popular language model is $n$-\emph{grams}, which has also been applied to model software code \cite{Hindle:2012:NS,Nguyen:2013:SSL}. This method truncates the history length to $n-1$ words, and in practice $n$ is often limited to a small integer (e.g., 2 to 5). Although $n$-\emph{grams} models are useful and intuitive in making use of repetitive sequential patterns in code, their context is restricted to a few code elements and is thus not sufficient in complex SE tasks. Recent work \cite{Tu:2014:LS} introduces a local cache model, which captures local regularities, combining with the global $n$-\emph{grams} model. Although this approach has successfully dealt with the localness of software code, it still suffers from the small context problem inherently in the $n$-\emph{grams} method.

Deep learning-based approaches offers a powerful alternative to $n$-\emph{grams} in representing software code. While deep feed-forward architectures such as convolutional neural networks (CNNs) are effective in capturing rich structural patterns in code \cite{mou2016convolutional}, recurrent architectures like Recurrent Neural Networks (RNNs) has the potential to capture a much longer context than $n$-\emph{grams} \cite{White:2015:TDL}. RNNs however suffer from the vanishing or exploding gradients \cite{bengio2013advances}, and thus make it very hard to train for long sequences. Consider trying to predict next code token in the code fragment: \texttt{String conferences = ["ICSE", "FSE", "ASE"]; ..... for (String conf :\underline{\space\space}}. Recent information suggest that the next code token is probably a set of Strings, but if we want to narrow down to a specific set, we need the context of \texttt{conferences} set, from further back. There could be a big gap between relevant information and the point where it is needed. Unfortunately, as that gap increases, RNNs will not be able to learn to connect the information.

In this paper, we propose a novel approach to build a language model for software code using Long Short-Term Memory (LSTM) \cite{hochreiter1997long}, a special kind of RNN that is capable of learning \emph{long-term dependencies}, i.e. remembering information from further back. LSTMs have demonstrated ground-breaking performance in many applications such as machine translation, video analysis, and speed recognition. Results from our preliminary evaluation have shown the effectiveness of LSTM, serving as concrete indication that LSTM is a promising model for software code. This code modeling component is part of our vision for DeepSoft \cite{DeepSoftFSE2016}, an \emph{end-to-end}, generic deep learning-based framework for modeling software and its development process. 

\section{Approach}

\begin{figure}[ht]
	\centering
	\includegraphics[width=\linewidth]{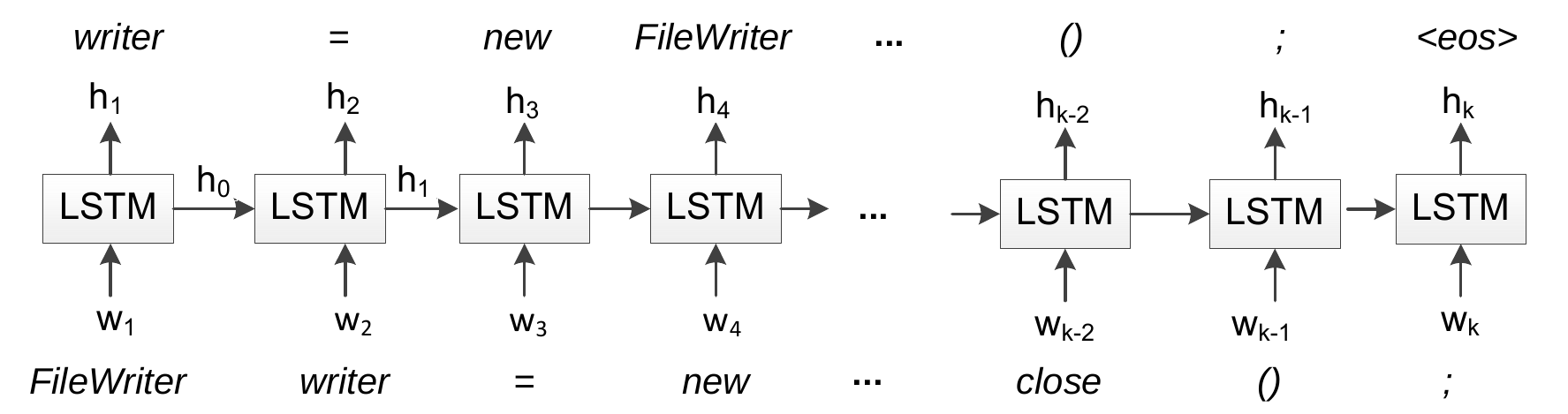}
	\caption{Long Short-Term Memory language model}
	\label{fig:DeepCode}
\end{figure}

Our LSTM language model of software code is a (special) recurrent neural network, which can be seen as multiple copies of the same network (see multiple LSTM units in Figure \ref{fig:DeepCode}), each passing information to a successor and thus allowing information to persist. Our language model reads code tokens in a sequence one by one, and estimates the probability of the next code token by the following steps. First, the current code token $w_t$ is mapped into a continuous space using lookup table $\mathcal{M}$ such that $\wb_{t}=\mathcal{M}(w)$ is a vector in $\mathbb{R}^{D}$. Vector $\wb_{t}$ then serves as an input to an LSTM unit. For example, each code token (e.g., \texttt{FileWriter}) in Listing \ref{example} is embedded into an input vector (e.g., $\wb_{1}$).

A standard RNN unit would then generate the hidden state, represented by vector $\hb_{t}\in\mathbb{R}^{K}$, according to the previous hidden state $\hb_{t-1}$ and the current input $\wb_{t}$.

\begin{equation}\label{eq:hidden-state}
\hb_{t}=\sigma\left(\bb+W_{tran}\hb_{t-1}+W_{in}\wb_{t}\right)
\end{equation}
where $\sigma$ is a nonlinear element-wise transform function, and $\bb$, $W_{tran}$ and $W_{in}$
are parameters. 

Finally, the next code token is predicted using:

\begin{equation}
P\left(w\mid\wb_{1:t-1}\right)=\frac{\exp\left(U_{w}^{\top}\hb_{t-1}\right)}{\sum_{u\in\mathscr{V}}\exp\left(U_{u}^{\top}\hb_{t-1}\right)}\label{eq:softmax}
\end{equation}
where $U_{w}\in\mathbb{R}^{K}$ is a free parameter.

The language model is then trained using many known sequences of code tokens $s=\langle w_{1},w_{2},...,w_{k}\rangle$ in a code corpus. Learning is typically done by minimizing the log-loss $-\log P(s)$ with respect to model parameters $\thetab$, which are 
$\left(\bb,W_{tran},W_{in},\mathcal{M},U\right)$:

\begin{equation}\label{eq:log-loss}
L(\thetab)=-\log P(w_{1})-\sum_{t=2}^{k}\log P\left(w_{t}\mid\wb_{1:t-1}\right)
\end{equation}

Learning involves computing the gradient of $L(\thetab)$ during the backpropagation phase, and parameters
are updated using a stochastic gradient descent (SGD). It means that parameters are updated after seeing only a small random subset of sequences. The critical problem with standard RNNs however lies here where the magnitude of weights in the transition matrix $W_{tran}$ (used in Equation~\ref{eq:hidden-state}) can have a strong effect on the learning process. This is because during the gradient backpropagation phase, the gradient signal can be multiplied a large number of times (as many as the number of code tokens in a sequence) by the transition matrix. If the weights in this matrix are small, it can result in the problem of \emph{vanishing gradients} where the gradient signal becomes so small that the learning gets very slow or even stopped. This makes standard RNNs not suitable for learning long-term dependencies. On the other hand, the large weights can lead to a large gradient signal (e.g. \emph{exploding gradient}), which can cause learning to diverge.

\begin{lstlisting}[caption=A motivating example, label=example]
  FileWriter writer = new FileWriter(file);
  writer.write(``This is an example'');
  int count = 0;
  System.out.prinltln(``Long gap'');
      ......

  writer.flush();
  writer.close();
\end{lstlisting}

\begin{figure}[ht]
	\centering
	\includegraphics[width=0.5\linewidth]{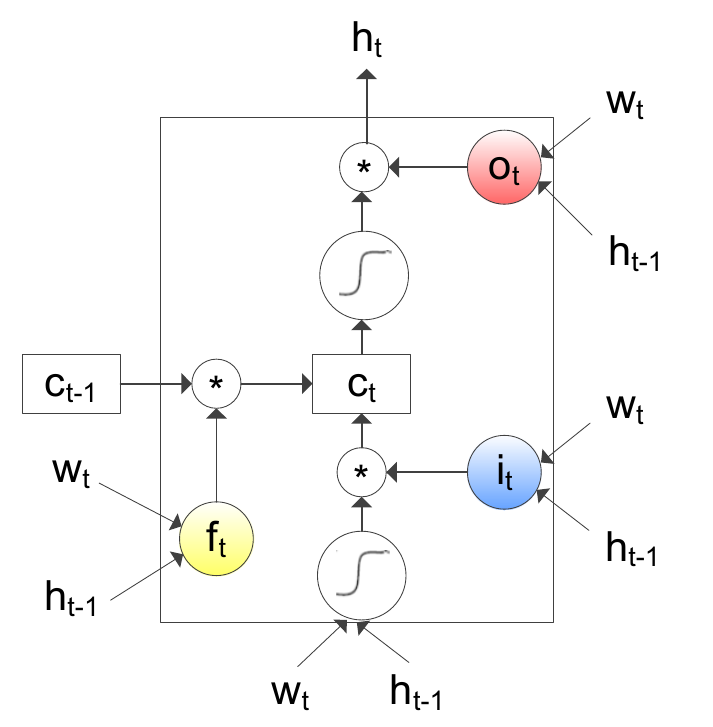}
	\caption{The internal structure of an LSTM for code processing}
	\label{fig:LSTMunit}
\end{figure}

The most powerful feature of our model (compared to the standard RNNs model) resides in the \emph{memory cell}  ($c_t$ in Figure \ref{fig:LSTMunit}) of an LSTM unit. This cell stores accumulated memory of the context, which is a mechanism to deal the vanish and exploding gradient problems. The amount of information flowing through the memory cell is controlled by three gates (an \emph{input gate}, a \emph{forget gate}, and an \emph{output gate}), each of which returns a value between 0 (i.e. complete blockage) and 1 (full passing through). All these gates are \emph{learnable}, i.e. being trained with the whole code corpus to maximize the predictive performance of the language model.

Let us now explain how an LSTM unit works. First, an LSTM unit decides how much information from the memory of previous context  (i.e. $c_{t-1}$) should be removed from the memory cell. This is controlled by the forget gate $\fb_{t}$, which looks at the the previous output state $h_{t-1}$ and the current code token $w_t$, and outputs a number between 0 and 1. A value of 1 indicates that all the past memory is preserved, while a value of 0 means ``completely forget everything''. In our example in Listing \ref{example}, the memory cell might keep information about the \texttt{writer} stream, so that it remembers this open stream needs closing later. When we see this stream being closed, we want to forget it.

The next step is updating the memory with new information obtained from the current code token $w_t$. In a similar manner to the forget gate, the input gate $\ib_{t}$ is used to control which new information will be stored in the memory. For example, if $w_t$ is the new open stream, we want to add it into the memory cell so that we remember closing it later. Finally, information stored in the memory cell will be used to produce an output $h_t$. The output gate $\ob_t$ looks at the current code token $w_t$ and the previous hidden state $h_{t-1}$, and determines which parts of the memory should be output. For example, since it just saw the token \texttt{new}, it might want to output information relevant (e.g. constructor forms) to the initialization of the object declared previously, in case that what is coming next. It is important to note here that LSTM computes the current hidden state based on not just only the current input $w_t$ and the previous hidden state $h_{t-1}$ (as done in standard RNNs) but also the current memory cell state $c_t$, which is \emph{linear} with the previous memory $c_{t-1}$. This is the key feature allowing our model to learn long-term dependencies in software code.

\subsection{Model training}

The bottleneck of our language model (as with any other RNN-based model) is the cost of normalization of the RHS of  Equation~(\ref{eq:softmax}) (i.e. the \emph{softmax} function on the gradient log), which scales with vocabulary size $N=\left|\mathscr{V}\right|$. In a typical software language model, the vocabulary  size can be very large (e.g., up to millions for large repositories). We have employed an efficient method called Noise Contrastive Estimation (NCE) \cite{gutmann2012noise}, which has been used successfully in natural language modeling (e.g., see \cite{chen2015recurrent}). Using NCE, the computation time now does not depend on the vocabulary size, but on a user-defined parameter $k$ that approximates the distribution in Equation~(\ref{eq:softmax}). In practice, $k$ is 100 or 200.


Two largest parameter matrices are the embedding table $\mathcal{M}\in\mathbb{R}^{D\times N}$ (which maps a code token into a continuous vector) and the prediction matrix $U\in\mathbb{R}^{D\times K}$ used in Equation~(\ref{eq:softmax}). For a large vocabulary, the number of parameters grows to millions, causing a potential overfitting. An effective solution is \emph{dropout} \cite{srivastava2014dropout}, where the elements of input embedding and output states are randomly set to zeros during training. During testing, a parameter averaging is used. In effect, dropout implicitly trains many models in parallel, and all of them share the same parameter set. The final model is an average of all models. Typically, the dropout rate is set at $0.5$. Another strategy to combat overfitting is early stopping, which we have employed for building our model. Here, we maintain an independent validation dataset to monitor the model performance during training, stop when the performance gets worse, and select the best performing model in the validation set.


\section{Evaluation}

We built a dataset of the ten Java projects (Ant, Batik, Cassandra, Eclipse-E4, Log4J, Lucene, Maven2, Maven3, Xalan-J, and Xerces) cloned from GitHub. These projects are the same as those  used in previous studies \cite{Nguyen:2013:SSL,Hindle:2012:NS} but we collected the most up-to-date revision of the code at the time, i.e. 2016-07-25. After removing comments and blank lines, the projects were lexically analyzed using JavaParser\footnote{\url{http://javaparser.org}} to produce token sequences. We then partitioned the data into mutually exclusive training, validation, and test sets. The training set was used to learn a useful language. After each training epoch, the learned model was evaluated on the validation set and its performance is used to assess convergence against hyperparameters (e.g. learning rate in gradient searches). Note that the validation set was \emph{not} used to learn any of the model's parameters. The best performing model in the validation set was chosen to be evaluated on the test set.

Following common practices (e.g., as done in \cite{White:2015:TDL}), we replaced integers, real numbers, exponential notation, hexadecimal numbers with a generic \emph{<num>} token, and replaced constant strings with a generic \emph{<str>} token. We also replaced less popular tokens (e.g. occurring only once in the corpus) and tokens which exist in the validation and test sets but do not exist in the training set with a special token \emph{<unk>}. Finally, we build a code corpus of 6,103,191 code tokens, with a vocabulary of 81,213 \emph{unique} tokens.

Each source file is parsed into a sequence of tokens. Each sequence is then split into sentences (i.e a subsequence of code tokens) of fixed length. We use 10K sentences for each case of training, validation and testing. Sentence length varies from 10 (e.g., a short statement), to 500 (e.g., a large file). Embedding dimensionality ($D$) varies from 20 to 500. For simplicity, the size of the memory (i.e. the size of $\cb_{t}$) is the same as the number of embedding dimensions, that is $D=K$. A fixed-size vocabulary is constructed based on top $N$ popular tokens, and rare tokens are assigned to \emph{<unk>}.

Our evaluation aims to compare the performance of our LSTM language model against the traditional RNNs model as used in previous work \cite{White:2015:TDL}. We also used \emph{perplexity} (as done in \cite{White:2015:TDL}), an intrinsic evaluation metric that estimates the average number of code tokens to select from at each point in a sequence. Specifically, perplexity is the inverse of average probability per code token, which is computed as
$\exp(-\sum_s\log P(s)/\#words)$, where $-\log P(s)$ is the log-loss $L(\thetab)$
computed in Equation~\ref{eq:log-loss}. A smaller perplexity implies a better language model.

Both our LSTM model and the simple RNN model were trained using an adaptive stochastic gradient descent method called RMSprop, which is known to work best for recurrent models. There are three hyperparameters: learning rate $\eta$, adaptation rate $\rho$ and smoothing factor $\epsilon$.
RMSprop is tuned to achieve best results for simple RNN ($\eta = 0.01$, $\rho = 0.9$ and $\epsilon = 10^{-8}$); and for LSTM ($\eta = 0.02$, $\rho = 0.99$ and $\epsilon = 10^{-7}$), respectively.

\begin{table}
\begin{centering}
\begin{tabular}{|c|c|c|c|c|}
\hline 
sent-len & embed-dim & RNN & LSTM & improv \%\tabularnewline
\hline 
\hline 
10 & \multirow{6}{*}{50} & 13.49 & 12.86 & \textbf{4.7}\tabularnewline
\cline{1-1} \cline{3-5} 
20 &  & 10.38 & 9.66 & \textbf{6.9}\tabularnewline
\cline{1-1} \cline{3-5} 
50 &  & 7.93 & 6.81 & \textbf{14.1}\tabularnewline
\cline{1-1} \cline{3-5} 
100 &  & 7.20 & 6.40 & \textbf{11.1}\tabularnewline
\cline{1-1} \cline{3-5} 
200 &  & 6.64 & 5.60 & \textbf{15.7}\tabularnewline
\cline{1-1} \cline{3-5} 
500 &  & 6.48 & 4.72 & \textbf{27.2}\tabularnewline
\hline 
\hline 
\multirow{4}{*}{100} & 20 & 7.96 & 7.11 & \textbf{10.7}\tabularnewline
\cline{2-5} 
 & 50 & 7.20 & 6.40 & \textbf{11.1}\tabularnewline
\cline{2-5} 
 & 100 & 7.23 & 5.72 & \textbf{20.9}\tabularnewline
\cline{2-5} 
 & 200 & 9.14 & 5.68 & \textbf{37.9}\tabularnewline
\hline 
\end{tabular}
\par\end{centering}
\caption{Perplexity on test data (the smaller the better). \label{tab:Perplexity-on-test}}

\end{table}

We conducted a number of experiments by varying the size of a sentence (\emph{sent-len}) and the number of embedding dimensions (\emph{embed-dim}). Table~\ref{tab:Perplexity-on-test} reports the perplexity on the test data for a vocab size of $N=1,000$, when fixing embedding dimensionality $D$ (top part) and when fixing sentence length (bottom part).

Overall, given a fixed embedding dimensionality, both simple RNN and LSTM models improve with more training data (whose size grows with sentence length), and the LSTM performs consistently better.  Importantly, the rate of improvement by LSTM also grows from 4.7\% for short sentences of length 10, to 27.2\% for long sentences of length 500, confirming the known fact that the LSTM handles long sequences better than the simple RNN. A similar pattern of improvement is also observed when fixing sentence length and varying dimensionality. It suggests that LSTM enjoys a better learning dynamic when the model is large.

\section{Discussion}

The promising results from our evaluation suggest that a good language model for software code can be built based on the powerful, deep learning-based LSTM architecture. Our next step involves conducting an extrinsic evaluation to measure the performance of our language model at a number of real SE tasks. For example, we will develop our language model into a \emph{code suggestion} engine. This engine will suggest a sequence of code tokens that fit most with the current context and is most likely to appear next. 

Our language model can also provide us with a deep, semantic representation for any arbitrary sequence of code tokens. This can be done by aggregating all the output vectors ($h_{1}$, ..., $h_k$) through a mechanism known as pooling. The simplest method is mean-pooling where the vector representing a code sequence is the sum of the output vectors of all the code token in the sequence divided by the number of tokens. This representation is in the form of a feature vector, which can be readily input to any classifier to learn many useful SE tasks such as defect prediction, checking for code duplication, and detecting vulnerable software components. As we have seen, this feature vector is learned automatically (in an unsupervised manner) from the raw code, removing us from the time-consuming task of manually designing features (which have been done in previous work).

The language model for software code is part of DeepSoft -- our generic, dynamic deep learning-based framework for modeling software and its development and evolution process. DeepSoft is a compositional architecture which has several layers modeling the progression of a software at four levels: source code, issue reports, releases, and project. Each layer employs chain of LSTM units  to capture the sequential nature of the corresponding data. We envision many applications of DeepSoft to SE problems ranging from requirements to maintenance such as risk prediction, effort estimation, issue resolution recommendation, code patch generation, release planning, and developer modeling.



\small
\bibliographystyle{abbrv}

\begin{thebibliography}{10}

\bibitem{bengio2013advances}
Y.~Bengio, N.~Boulanger-Lewandowski, and R.~Pascanu.
\newblock Advances in optimizing recurrent networks.
\newblock In {\em 2013 IEEE International Conference on Acoustics, Speech and
  Signal Processing}, pages 8624--8628. IEEE, 2013.

\bibitem{chen2015recurrent}
X.~Chen, X.~Liu, M.~J. Gales, and P.~C. Woodland.
\newblock Recurrent neural network language model training with noise
  contrastive estimation for speech recognition.
\newblock In {\em 2015 IEEE International Conference on Acoustics, Speech and
  Signal Processing (ICASSP)}, pages 5411--5415. IEEE, 2015.

\bibitem{DeepSoftFSE2016}
H.~K. Dam, T.~Tran, J.~Grundy, and A.~Ghose.
\newblock {DeepSoft: A vision for a deep model of software}.
\newblock In {\em Proceedings of the 24th ACM SIGSOFT International Symposium
  on Foundations of Software Engineering}, FSE '16. ACM, To Appear., 2016.

\bibitem{Gabel:2010:SUS}
M.~Gabel and Z.~Su.
\newblock A study of the uniqueness of source code.
\newblock In {\em Proceedings of the Eighteenth ACM SIGSOFT International
  Symposium on Foundations of Software Engineering}, FSE '10, pages 147--156,
  New York, NY, USA, 2010. ACM.

\bibitem{gutmann2012noise}
M.~U. Gutmann and A.~Hyv{\"a}rinen.
\newblock Noise-contrastive estimation of unnormalized statistical models, with
  applications to natural image statistics.
\newblock {\em Journal of Machine Learning Research}, 13(Feb):307--361, 2012.

\bibitem{Hindle:2012:NS}
A.~Hindle, E.~T. Barr, Z.~Su, M.~Gabel, and P.~Devanbu.
\newblock On the naturalness of software.
\newblock In {\em Proceedings of the 34th International Conference on Software
  Engineering}, ICSE '12, pages 837--847, Piscataway, NJ, USA, 2012. IEEE
  Press.

\bibitem{hochreiter1997long}
S.~Hochreiter and J.~Schmidhuber.
\newblock Long short-term memory.
\newblock {\em Neural computation}, 9(8):1735--1780, 1997.

\bibitem{mou2016convolutional}
L.~Mou, G.~Li, L.~Zhang, T.~Wang, and Z.~Jin.
\newblock Convolutional neural networks over tree structures for programming
  language processing.
\newblock In {\em Proceedings of the Thirtieth AAAI Conference on Artificial
  Intelligence}, 2016.

\bibitem{Nguyen:2013:SSL}
T.~T. Nguyen, A.~T. Nguyen, H.~A. Nguyen, and T.~N. Nguyen.
\newblock A statistical semantic language model for source code.
\newblock In {\em Proceedings of the 2013 9th Joint Meeting on Foundations of
  Software Engineering}, ESEC/FSE 2013, pages 532--542, New York, NY, USA,
  2013. ACM.

\bibitem{srivastava2014dropout}
N.~Srivastava, G.~Hinton, A.~Krizhevsky, I.~Sutskever, and R.~Salakhutdinov.
\newblock Dropout: A simple way to prevent neural networks from overfitting.
\newblock {\em Journal of Machine Learning Research}, 15:1929--1958, 2014.

\bibitem{Tu:2014:LS}
Z.~Tu, Z.~Su, and P.~Devanbu.
\newblock On the localness of software.
\newblock In {\em Proceedings of the 22Nd ACM SIGSOFT International Symposium
  on Foundations of Software Engineering}, FSE 2014, pages 269--280, New York,
  NY, USA, 2014. ACM.

\bibitem{White:2015:TDL}
M.~White, C.~Vendome, M.~Linares-V\'{a}squez, and D.~Poshyvanyk.
\newblock Toward deep learning software repositories.
\newblock In {\em Proceedings of the 12th Working Conference on Mining Software
  Repositories}, MSR '15, pages 334--345, Piscataway, NJ, USA, 2015. IEEE
  Press.

\end{thebibliography}

\end{document}